\documentclass[utf8]{frontiersFPHY} 
\setcitestyle{square} 

\usepackage{url,hyperref,lineno,microtype,subcaption}
\usepackage[onehalfspacing]{setspace}

\usepackage[page,title]{appendix}
\usepackage{url}
\usepackage{graphicx}

\usepackage{natbib}
\usepackage{graphicx}
\usepackage{xspace}
\newcommand{\unit}[1]{\ensuremath{\text{\,#1}}\xspace}
\newcommand{\emtkid}{\emph{EmTrackMichelId}\xspace}
\newcommand{\ptidalg}{\emph{PointIdAlg}\xspace}
\newcommand{\uboone}{MicroBooNE\xspace}
\newcommand{\nova}{NO$\nu$A\xspace}
\newcommand{\tcpu}{\ensuremath{t_\text{CPU}}\xspace}
\newcommand{\tgpu}{\ensuremath{t_\text{GPU}}\xspace}
\newcommand{\ttransmit}{\ensuremath{t_\text{transmit}}\xspace}
\newcommand{\ttravel}{\ensuremath{t_\text{travel}}\xspace}
\newcommand{\tlatency}{\ensuremath{t_\text{latency}}\xspace}
\newcommand{\tideal}{\ensuremath{t_\text{ideal}}\xspace}
\newcommand{\ncpu}{\ensuremath{N_\text{CPU}}\xspace}
\newcommand{\ngpu}{\ensuremath{N_\text{GPU}}\xspace}
\newcommand{\nreq}{\ensuremath{N_\text{request}}\xspace}
\newcommand{\treq}{\ensuremath{t_\text{request}}\xspace}
\newcommand{\tsonic}{\ensuremath{t_\text{SONIC}}\xspace}
\newcommand{\tpreproc}{\ensuremath{t_\text{preprocess}}\xspace}

\def\firstAuthorLast{Wang {et~al.}} 
\def\Authors{Michael~Wang\,$^{1,*}$, Tingjun~Yang\,$^{1}$,  Maria~Acosta~Flechas\,$^{1}$, Philip Harris\,$^{2}$, Benjamin Hawks\,$^{1}$, Burt Holzman\,$^{1}$, Kyle Knoepfel\,$^{1}$, Jeffrey Krupa\,$^{2}$, Kevin Pedro\,$^{1}$, Nhan Tran\,$^{1,3}$}
\def\Address{$^{1}$ Fermi National Accelerator Laboratory, Batavia, IL 60510, USA \\
$^{2}$ Massachusetts Institute of Technology, Cambridge, MA 02139, USA \\
$^{3}$ Northwestern University, Evanston, IL 60208, USA
}
\def\corrAuthor{Michael Wang}

\def\corrEmail{mwang@fnal.gov}

\begin{document}

\onecolumn
\firstpage{1}

\title[GPU-accelerated ML inference aaS for computing in neutrino experiments]{GPU-accelerated machine learning inference as a service for computing in neutrino experiments} 

\author[\firstAuthorLast ]{\Authors} 
\address{} 
\correspondance{} 

\extraAuth{}

\maketitle

\begin{abstract}

Machine learning algorithms are becoming increasingly prevalent and performant in the reconstruction of events in accelerator-based neutrino experiments. These sophisticated algorithms can be computationally expensive. At the same time, the data volumes of such experiments are rapidly increasing. The demand to process billions of neutrino events with many machine learning algorithm inferences creates a computing challenge. We explore a computing model in which heterogeneous computing with GPU coprocessors is made available as a web service. The coprocessors can be efficiently and elastically deployed to provide the right amount of computing for a given processing task. With our approach,  Services for Optimized Network Inference on Coprocessors (SONIC), we integrate GPU acceleration specifically for the ProtoDUNE-SP reconstruction chain without disrupting the native computing workflow. With our integrated framework, we accelerate the most time-consuming task, track and particle shower hit identification, by a factor of 17. This results in a factor of 2.7 reduction in the total processing time when compared with CPU-only production. For this particular task, only 1 GPU is required for every 68 CPU threads, providing a cost-effective solution.

\end{abstract}

\section{Introduction}
Fundamental particle physics has pushed the boundaries of computing for decades.  
As detectors have become more sophisticated and granular, particle beams more intense, and data sets larger, the biggest fundamental physics experiments in the world have been confronted with massive computing challenges.

The Deep Underground Neutrino Experiment (DUNE)~\cite{Abi:2020wmh}, the future flagship neutrino experiment based at Fermi National Accelerator Laboratory (Fermilab), will conduct a rich program in neutrino and underground physics, including determination of the neutrino mass hierarchy~\cite{Qian:2015waa} and measurements of CP violation~\cite{Nunokawa:2007qh} in neutrino mixing using a long baseline accelerator-based neutrino beam, detection and measurements of atmospheric and solar neutrinos~\cite{Capozzi:2018dat}, searches for supernova-burst neutrinos~\cite{Scholberg:2012id} and other neutrino bursts from astronomical sources, and searches for physics at the grand unification scale via proton decay~\cite{Kudryavtsev:2016ybl}.

The detectors will consist of four modules, of which at least three are planned to be $10\unit{kton}$ Liquid Argon Time Projection Chambers (LArTPCs). Charged particles produced from neutrino or other particle interactions will travel through and ionize the argon, with ionization electrons drifted over many meters in a strong electric field and detected on planes of sensing wires or printed-circuit-board charge collectors. The result is essentially a high-definition image of a neutrino interaction, which naturally lends itself to applications of machine learning (ML) techniques designed for image classification, object detection, and semantic segmentation. ML can also aid in other important applications, like noise reduction and anomaly or region-of-interest detection.

Due to the number of channels and long readout times of the detectors, the data volume produced by the detectors will be very large: uncompressed continuous readout of a single module will be nearly $1.5\unit{TB}$ per second. Because that amount of data is impossible to collect and store, let alone process, and because most of that data will not contain interactions of interest, a real-time data selection scheme must be employed to identify data containing neutrino interactions. With a limit on total bandwidth of $30\unit{PB}$ of data per year for all DUNE modules, that data selection scheme, with accompanying compression, must effectively reduce the data rate by a factor of $10^6$.

In addition to applications in real-time data selection, accelerated ML inference that can scale to process large data volumes will be important for offline reconstruction and selection of neutrino interactions. The individual events are expected to have a size on the order of a few gigabytes, and extended readout events (associated, for example, with supernova burst events) may be significantly larger, up to $100\unit{TB}$ per module. It will be a challenge to efficiently analyze that data without an evolution of the computing models and technology that can handle data retrieval, transport, parallel processing, and storage in a cohesive manner. Moreover, similar computing challenges exist for a wide range of existing neutrino experiments such as \uboone~\cite{Acciarri:2016smi} and \nova~\cite{Ayres:2007tu}. 

In this paper, we focus on the acceleration of the inference of deep ML models as a solution for processing large amounts of data in the ProtoDUNE single-phase apparatus (ProtoDUNE-SP)~\cite{Abi:2020mwi} reconstruction workflow. 
For ProtoDUNE-SP, ML inference is the most computationally intensive part of the full event processing chain and is run repeatedly on hundreds of millions of events. A growing trend to improve computing power has been the development of hardware that is dedicated to accelerating certain kinds of computations. Pairing a specialized coprocessor with a traditional CPU, referred to as heterogeneous computing, greatly improves performance. These specialized coprocessors utilize natural parallelization and provide higher data throughput. In this study, the coprocessors employed are graphics processing units (GPUs); however, the approach can accommodate multiple types of coprocessors in the same workflow. ML algorithms, and in particular deep neural networks, are a driver of this computing architecture revolution.

For optimal integration of GPUs into the neutrino event processing workflow, we deploy them ``as a service.'' The specific approach is called Services for Optimized Network Inference on Coprocessors (SONIC)~\cite{Duarte:2019fta,SonicSW,Krupa:2020bwg}, which employs a client-server model. The primary processing job, including the clients, runs on the CPU, as is typically done in particle physics, and the ML model inference is performed on a GPU server. This can be contrasted with a more traditional model with a GPU directly connected to each CPU node. The SONIC approach allows a more flexible computing architecture for accelerating particle physics computing workflows, providing the optimal number of heterogeneous computing resources for a given task.

The rest of this paper is organized as follows. We first discuss the related works that motivated and informed this study. In Section~\ref{sec:setup}, we describe the tasks for ProtoDUNE-SP event processing and the specific reconstruction task for which an ML algorithm has been developed. We detail how the GPU coprocessors are integrated into the neutrino software framework as a service on the client side and how we set up and scale out GPU resources in the cloud. In Section~\ref{sec:results}, we present the results, which include single job and batch job multi-CPU/GPU latency and throughput measurements. Finally, in Section~\ref{sec:outlook}, we summarize the study and discuss further applications and future work.

\vspace{0.5cm}
\subsection*{Related Work}

Inference as a service was first employed for particle physics in Ref.~\cite{Duarte:2019fta}. This initial study utilized custom field programmable gate arrays (FPGAs) manufactured by Intel Altera and provided through the Microsoft Brainwave platform~\cite{configurable-cloud-acceleration}. These FPGAs achieved low-latency, high-throughput inference for large convolutional neural networks such as ResNet-50~\cite{DBLP:journals/corr/HeZRS15} using single-image batches. This acceleration of event processing was demonstrated for the Compact Muon Solenoid (CMS) experiment at the Large Hadron Collider (LHC), using a simplified workflow focused on inference with small batch sizes. Our study with GPUs for neutrino experiments focuses on large batch size inferences. GPUs are used in elements of the simulation of events in the IceCube Neutrino Observatory~\cite{Halzen:2010yj}; recently, a burst for the elements that run on GPUs was deployed at large scale~\cite{Sfiligoi_2020}. The ALICE experiment at the LHC is planning to use GPUs for real-time processing and data compression of their Time Project Chamber subdetectors~\cite{Rohr:2020fnu}.  The LHCb experiment at the LHC is considering using GPUs for the first level of their trigger system~\cite{Aaij:2019zbu}.

Modern deep ML algorithms have been embraced by the neutrino reconstruction community because popular computer vision and image processing techniques are highly compatible with the neutrino reconstruction task and the detectors that collect the data. \nova has applied a custom convolutional neural network (CNN), inspired by GoogLeNet~\cite{2014arXiv1409.4842S}, to the classification of neutrino interactions for their segmented liquid scintillator-based detector~\cite{novacvn}. \uboone, which uses a LArTPC detector, has conducted an extensive study of various CNN architectures and demonstrated their effectiveness in classifying and localizing single particles in a single wire plane, classifying neutrino events and localizing neutrino interactions in a single plane, and classifying neutrino events using all three planes~\cite{uboonecnn1}. In addition, \uboone has applied a class of CNNs, known as semantic segmentation networks, to 2D images formed from real data acquired from the LArTPC collection plane, in order to classify each pixel as being associated with an EM particle, other type of particle, or background~\cite{uboonecnn2}. DUNE, which will also use LArTPC detectors, has implemented a CNN based on the SE-ResNet~\cite{2017arXiv170901507H} architecture to classify neutrino interactions in simulated DUNE far detector events~\cite{dunecvn}.  Lastly, a recent effort has successfully demonstrated an extension of the 2D pixel-level semantic segmentation network from \uboone to three dimensions, using submanifold sparse convolutional networks~\cite{2017arXiv170601307G,kazusscn}.


\section{Setup and methodology}
\label{sec:setup}
In this study, we focus on a specific computing workflow, the ProtoDUNE-SP reconstruction chain, to demonstrate the power and flexibility of the SONIC approach. ProtoDUNE-SP, assembled and tested at the CERN Neutrino Platform (the NP04 experiment at CERN)~\cite{Pietropaolo:2017jlh}, is designed to act as a test bed and full-scale prototype for the elements of the first far detector module of DUNE. It is currently the largest LArTPC ever constructed and is vital to develop the technology required for DUNE. This includes the reconstruction algorithms that will extract physics objects from the data obtained using LArTPC detectors, as well as the associated computing workflows.

In this section, we will first describe the ProtoDUNE-SP reconstruction workflow and the ML model that is the current computing bottleneck. We will then describe the SONIC approach and how it was integrated into the LArTPC reconstruction software framework. Finally, we will describe how this approach can be scaled up to handle even larger workflows with heterogeneous computing.

\vspace{0.5cm}
\subsection{ProtoDUNE-SP reconstruction}\label{sec:workload}

The workflow used in this paper is the full offline reconstruction chain for the ProtoDUNE-SP detector, which is a good representative of event reconstruction in present and future accelerator-based neutrino experiments. In each event, ionizing particles pass through the liquid argon, emitting scintillation light that is recorded by photodetectors. The corresponding pulses are reconstructed as optical hits. These hits are grouped into flashes from which various parameters are determined, including time of arrival, spatial characteristics, and number of photoelectrons detected.

\begin{figure}[tbh]
\centering
\includegraphics[width=\textwidth]{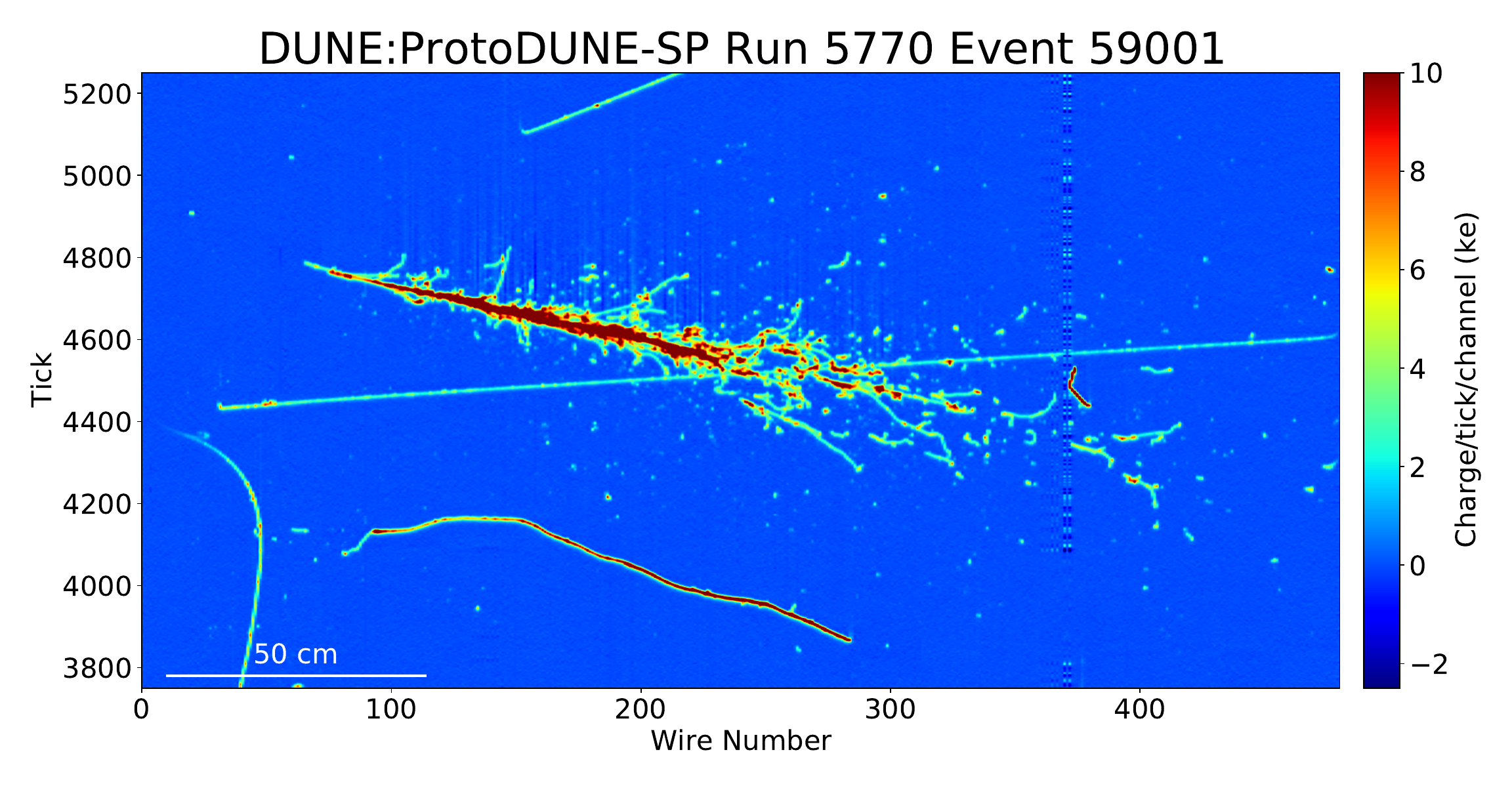}
\caption{A 6\unit{GeV/$c$} electron event in the ProtoDUNE detector. The x axis shows the wire number. The y axis shows the time tick in the unit of 0.5\unit{$\mu$s}. The color scale represents the charge deposition.}
\label{fig:evd}
\end{figure}

After the optical reconstruction stage, the workflow proceeds to the reconstruction of LArTPC wire hits. Figure~\ref{fig:evd} shows a 6\unit{GeV/$c$} electron event in the ProtoDUNE detector.

The reconstruction begins by applying a deconvolution procedure to recover the original waveforms by disentangling the effects of electronics and field responses after noise mitigation. The deconvolved waveforms are then used to find and reconstruct wire hits, providing information such as time and collected charge. Once the wire hits have been reconstructed, the 2D information provided by the hits in each plane is combined with that from the other planes in order to reconstruct 3D space points. This information is primarily used to resolve ambiguities caused by the induction wires in one plane wrapping around into another plane.

The disambiguated collection of reconstructed 2D hits is then fed into the next stage, which consists of a modular set of algorithms provided by the Pandora software development kit~\cite{Marshall:2013bda}. This stage finds the high-level objects associated with particles, like tracks, showers, and vertices, and assembles them into a hierarchy of parent-daughter nodes that ultimately point back to the candidate neutrino interaction.

The final module in the chain, \emtkid, is an ML algorithm that classifies reconstructed wire hits as being track-like, shower-like, or Michel electron-like~\cite{Michel_1950}. This algorithm begins by constructing $48\times48$ pixel images whose two dimensions are the time $t$ and the wire number $w$ in the plane. These images, called patches, are centered on the peak time and wire number of the reconstructed hit being classified. The value of each pixel corresponds to the measured charge deposition in the deconvolved waveforms for the wire number and time interval associated with the row and column, respectively, of the pixel.  Inference is performed on these patches using a convolutional neural network. Importantly, over the entire ProtoDUNE-SP detector, there are many $48\times48$ patches to be classified, such that a typical event may have ${\sim}55,000$ patches to process. Because of the way the data is processed per wire plane, those ${\sim}55,000$ patches are processed in batches with average sizes of either 235 or 1693. We will explore the performance implications of this choice in the next section. 

\vspace{0.5cm}
\subsection{CNN model for track identification}

\begin{figure}[tbh]
    \centering
    \includegraphics[width=0.7\textwidth]{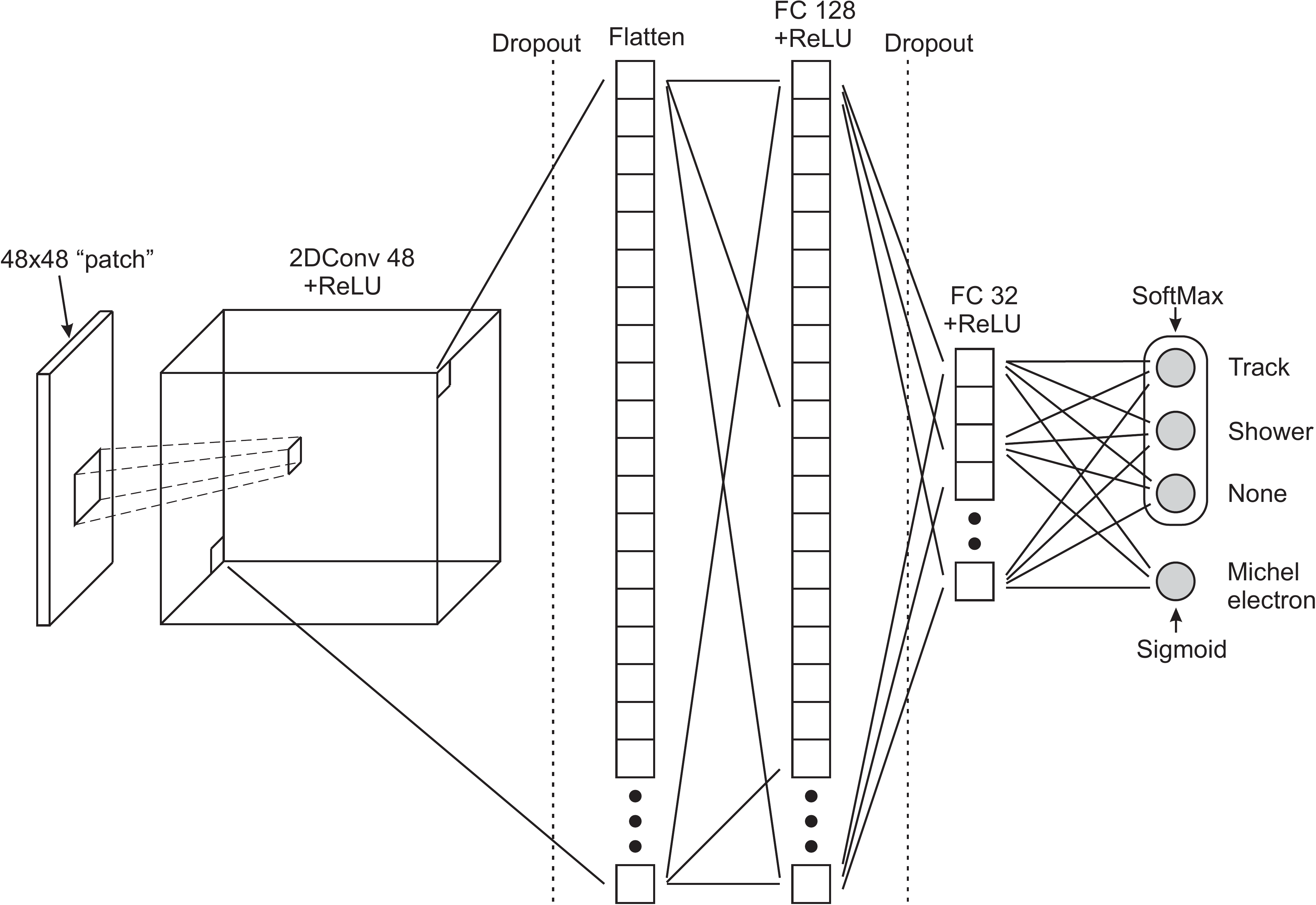}
    \caption{Architecture of the neural network used by the \emph{EmTrackMichelId} module in the ProtoDUNE-SP reconstruction chain, a convolutional (2DConv) layer is flattened to two fully connected layers.}
    \label{fig:model}
\end{figure}

The neural network employed by the \emtkid module of the ProtoDUNE-SP reconstruction chain consists of a 2D convolutional layer followed by two fully connected (FC) layers. The convolutional layer takes each of the $48\times48$ pixel patches described in Section~\ref{sec:workload} and applies 48 separate $5\times5$ convolutions to it, using stride lengths of 1, to produce 48 corresponding $44\times44$ pixel feature maps. These feature maps are then fed into the first fully connected (FC) layer consisting of 128 neurons, which is, in turn, connected to the second FC layer of 32 neurons. Rectified linear unit (ReLU) activation functions are applied after the convolutional layer and each of the two FC layers. Dropout layers are implemented between the convolutional layer and the first FC layer and between the two FC layers to help prevent overfitting. The second FC layer splits into two separate branches. The first branch terminates in three outputs that are constrained to a sum of one by a softmax activation function, and the second branch terminates in a single output with a sigmoid activation function that limits its value within a range of 0 to 1. The output of the network is split in this way due to the overlap of the shower and Michel electron classes. The second branch helps improve the efficiency of tagging hits produced by the Michel electrons, which are due to low energy shower activity. The total number of trainable parameters in this model is 11,900,420.  

\vspace{0.5cm}
\subsection{GPU inference as a service with LArSoft}

ProtoDUNE-SP reconstruction code is based on the LArSoft C++ software framework~\cite{Snider:2017wjd}, which provides a common set of tools shared by many LArTPC-based neutrino experiments. Within this framework, \emtkid, which is described in Section~\ref{sec:workload}, is a LArSoft ``module'' that makes use of the \ptidalg ``algorithm.''  \emtkid passes the wire number and peak time associated with a hit to \ptidalg, which constructs the patch and performs the inference task to classify it.

In this study, we follow the SONIC approach that is also in development for other particle physics applications. 
It is a client-server model, in which the coprocessor hardware used to accelerate the neural network inference is separate from the CPU client and accessed as a (web) service.  The neural network inputs are sent via TCP/IP network communication to the GPU.
In this case, a synchronous, blocking call is used. This means that the thread makes the web service request and then waits for the response from the server side, only proceeding once the server sends back the network output. In ProtoDUNE-SP, the CPU usage of the workflow, described in Section~\ref{sec:workload}, is dominated by the neural network inference. Therefore, a significant increase in throughput can still be achieved despite including the latency from the remote call while the CPU waits for the remote inference. An asynchronous, non-blocking call would be slightly more efficient, as it would allow the CPU to continue with other work while the remote call was ongoing. However, this would require significant development in LArSoft for applications of task-based multithreading, as described in Ref.~\cite{makortelCHEP2019}.

\begin{figure}[tbh]
    \centering
    \includegraphics[width=0.95\textwidth]{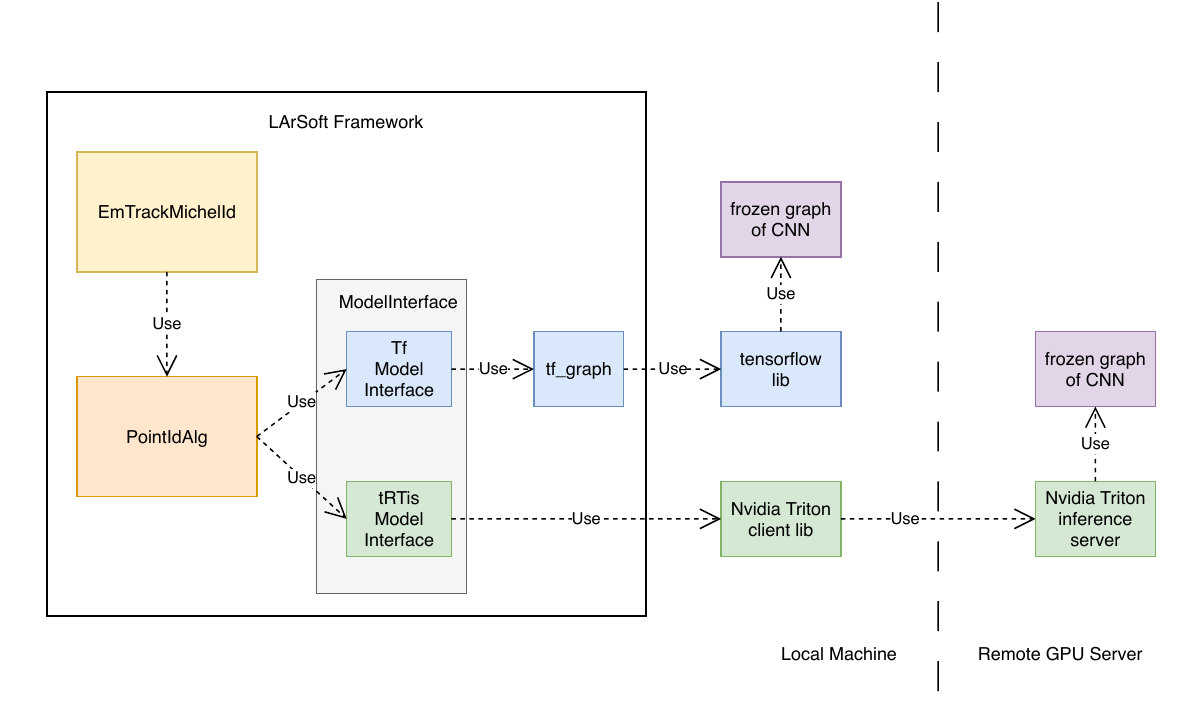}
    \caption{The interaction between \emph{EmTrackMichelId}, \emph{PointIdAlg}, and \emph{ModelInterface}, described in the text, is depicted in the figure above. CPU-only inference is provided by \emph{TfModelInterface}, while GPU-accelerated inference, via the GPUaaS approach, is provided by \emph{tRTisModelInterface}.}
\label{fig:gpuaasinlarsoft}
\end{figure}

The \emph{ModelInterface} class in LArSoft is used by \ptidalg to access the underlying ML model that performs the inference.  Previously, this inference was performed locally on a CPU using the \emph{TfModelInterface} subclass of \emph{ModelInterface}.
In this work, the functionality to access the GPU as a service was realized by implementing the C++ client interface, provided by the Nvidia Triton inference server~\cite{Triton}, in a new \emph{ModelInterface} subclass called \emph{tRTisModelInterface}. In this new subclass, the patches constructed by \ptidalg are put into the proper format and transmitted to the GPU server for processing, while a blocking operation ensues until inference results are received from the server. Communication between the client and server is achieved through remote procedure calls based on gRPC~\cite{gRPC}. The desired model interface subclass to use is selected and its parameters specified at run time by the user through a FHiCL~\cite{fhicl} configuration file.  The code is found in the \emph{LArSoft/larrecodnn} package~\cite{larrecodnn}. On the server side, we deploy NVidia T4 GPUs targeting data center acceleration. Figure \ref{fig:gpuaasinlarsoft} illustrates how \emph{EmTrackMichelId}, \emph{PointIdAlg}, and \emph{ModelInterface} interact with each other.  It also shows how each of the two available subclasses, \emph{TfModelInterface} and \emph{tRTisModelInterface}, access the underlying model.

\begin{figure}[tbh]
    \centering
    \includegraphics[width=0.95\textwidth]{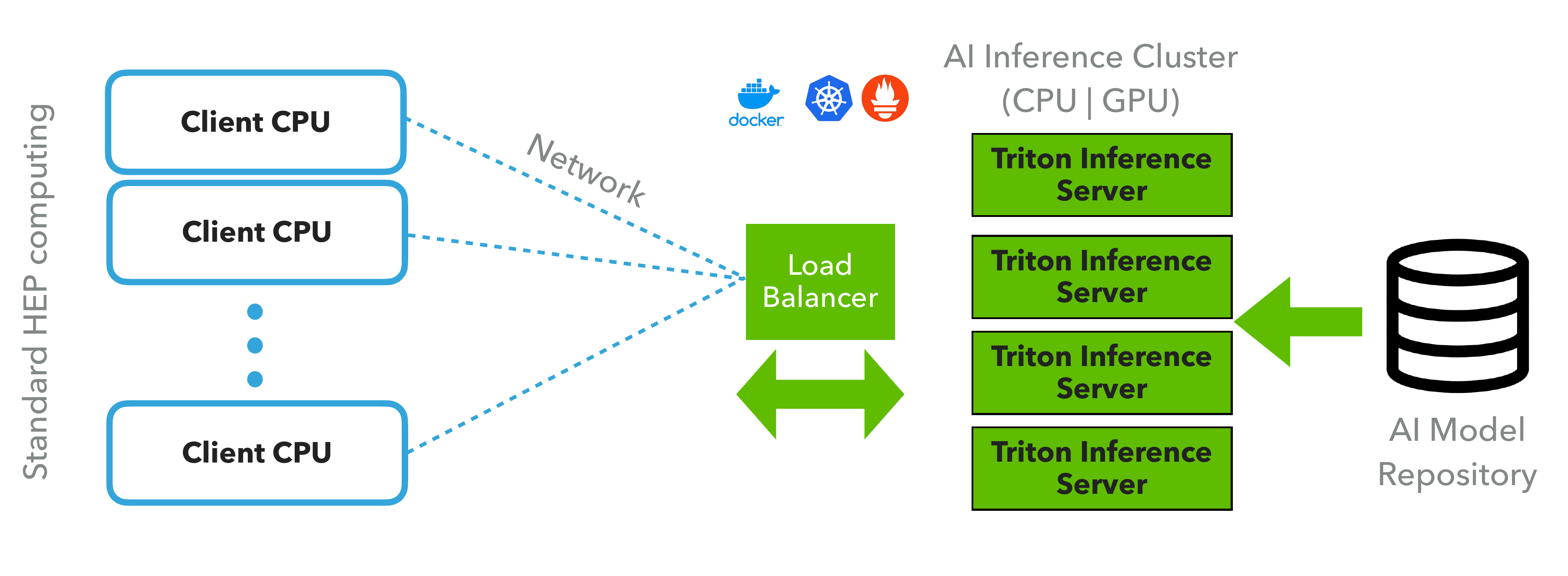}
    \caption{Depiction of the client-server model using Triton where multiple CPU processes on the client side are accessing the AI model on the server side. }
    \label{fig:aas}
\end{figure}

This flexible approach has several advantages:
\begin{itemize}
    \item Rather than requiring one coprocessor per CPU with a direct connection over PCIe, many worker nodes can send requests to a single GPU, as depicted in Fig.~\ref{fig:aas}. This allows heterogeneous resources to be allocated and re-allocated based on demand and task, providing significant flexibility and potential cost reduction. The CPU-GPU system can be ``right-sized'' to the task at hand, and with modern server orchestration tools, described in the next section, it can elastically deploy coprocessors.
    \item There is a reduced dependency on open-source ML frameworks in the experimental code base. Otherwise, the experiment would be required to integrate and support separate C++ APIs for every framework in use.
    \item In addition to coprocessor resource scaling flexibility, this approach allows the event processing to use multiple types of heterogeneous computing hardware in the same job, making it possible to match the computing hardware to the ML algorithm. The system could, for example, use both FPGAs and GPUs servers to accelerate different tasks in the same workflow.
\end{itemize}

There are also challenges to implementing a computing model that accesses coprocessors as a service. Orchestration of the client-server model can be more complicated, though we find that this is facilitated with modern tools like the Triton inference server and Kubernetes. In Section~\ref{sec:outlook}, we will discuss future plans to demonstrate production at full scale. Networking from client to server incurs additional latency, which may lead to bottlenecks from limited bandwidth. For this particular application, we account for and measure the additional latency from network bandwidth, and it is a small, though non-negligible, contribution to the overall processing.

The Triton software also handles load balancing for servers that provide multiple GPUs, further increasing the flexibility of the server. In addition, the Triton server can host multiple models from various ML frameworks. One particularly powerful feature of the Triton inference server is dynamic batching, which combines multiple requests into optimally-sized batches in order to perform inference as efficiently as possible for the task at hand. This effectively enables simultaneous processing of multiple events without any changes to the experiment software framework, which assumes one event is processed at a time.

\vspace{0.5cm}
\subsection{Kubernetes scale out}
\label{sec:k8s}
We performed tests on many different combinations of computing hardware, which provided a deeper understanding of networking limitations within both Google Cloud and on-premises data centers. Even though the Triton Inference Server does not consume significant CPU power, the number of CPU cores provisioned for the node did have an impact on the maximum ingress bandwidth achieved in the early tests.

\begin{figure}[tbh]
    \centering
    \includegraphics[width=0.8\textwidth]{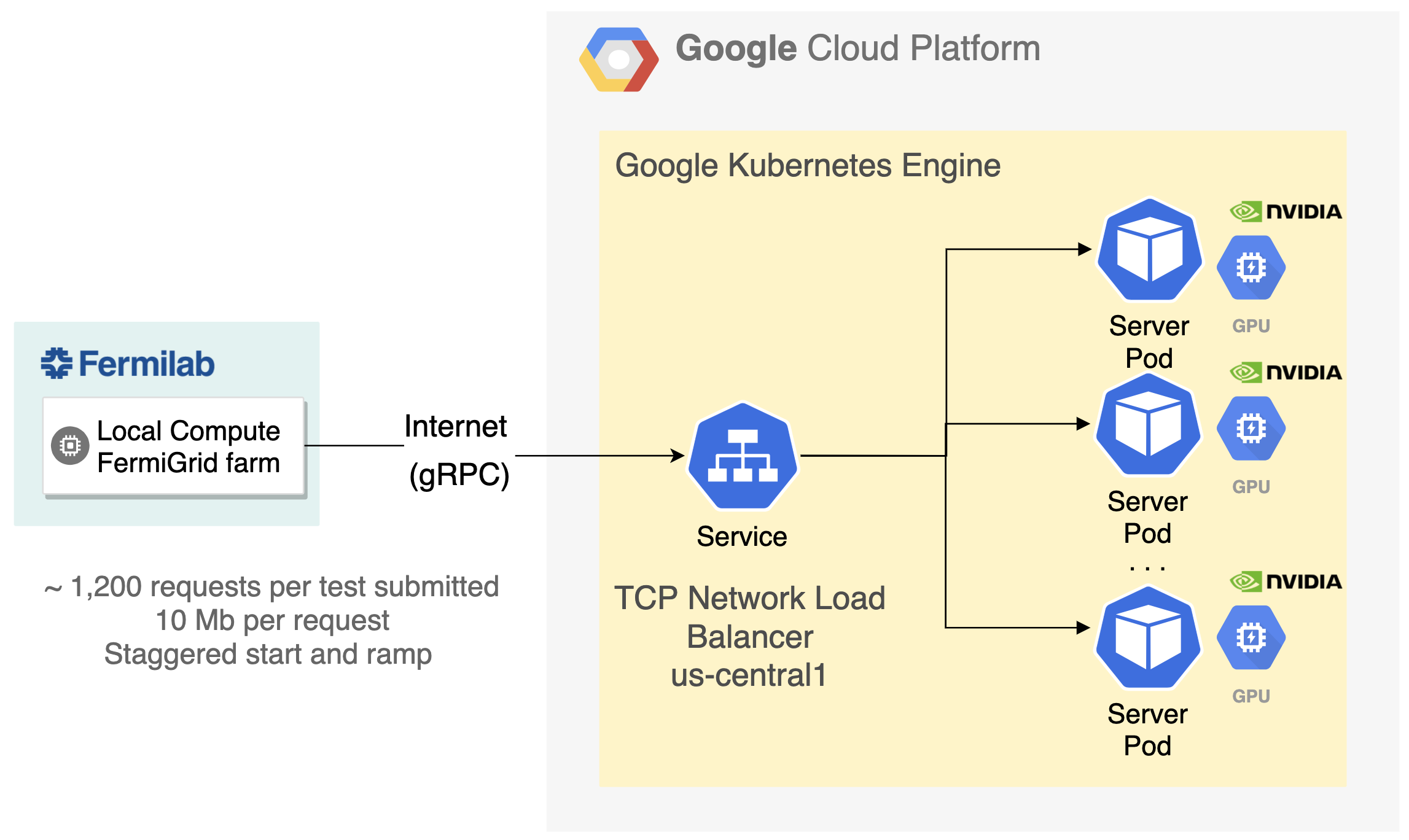}
    \caption{The Google Kubernetes Engine setup which demonstrates how the Local Compute FermiGrid farm communicates with the GPU server and how the server is orchestrated through Kubernetes.}
    \label{fig:gke}
\end{figure}

To scale the NVidia T4 GPU throughput flexibly, we deployed a Google Kubernetes Engine (GKE) cluster for server-side workloads.  The cluster is deployed in the US-Central data center, which is located in Iowa; this impacts the data travel latency. The cluster was configured using a Deployment and ReplicaSet. These are Kubernetes artifacts for application deployment, management and control. They hold resource requests, container definitions, persistent volumes, and other information describing the desired state of the containerized infrastructure. 
Additionally, a load-balancing service to distribute incoming network traffic among the Pods was deployed. We implemented Prometheus-based monitoring, which provided insight into three aspects: system metrics for the underlying virtual machine, Kubernetes metrics on the overall health and state of the cluster, and inference-specific metrics gathered from the Triton Inference Server via a built-in Prometheus publisher. All metrics were visualized through a Grafana instance, also deployed within the same cluster. The setup is depicted in Fig.~\ref{fig:gke}.

A Pod is a group of one or more containers with shared storage and network, and a specification for how to run the containers. A Pod’s contents are always co-located and co-scheduled, and run in a shared context within Kubernetes Nodes~\cite{k8sDocs}. 
We kept the Pod to Node ratio at 1:1 throughout the studies, with each Pod running an instance of the Triton Inference Server (v20.02-py3) from the Nvidia Docker repository. The Pod hardware requests aim to maximize the use of allocatable virtual CPU (vCPU) and memory and to use all GPUs available to the container. 

In this scenario, it can be naively assumed that a small instance n1-standard-2 with 2 vCPUs, $7.5\unit{GB}$ of memory, and different GPU configurations (1, 2, 4, or 8) would be able to handle the workload, which would be distributed evenly on GPUs. After performing several tests, we found that horizontal scaling would allow us to increase our ingress bandwidth, because Google Cloud imposes a hard limit on network bandwidth of $2\unit{Gbps}$ per vCPU, up to a theoretical maximum of $16\unit{Gbps}$ for each virtual machine~\cite{gcpNetDocs}. 

Given these parameters, we found that the ideal setup for optimizing ingress bandwidth was to provision multiple Pods on 16-vCPU machines with fewer GPUs per Pod. For GPU-intensive tests, we took advantage of having a single point of entry, with Kubernetes balancing the load and provisioning multiple identical Pods behind the scenes, with the total GPU requirement defined as the sum of the GPUs attached to each Pod.

\section{Results}
\label{sec:results}
Using the setup described in the previous section to deploy GPUs as a service (GPUaaS) to accelerate machine learning inference, we measure the performance and compare against the default CPU-only workflow in ProtoDUNE-SP.

First, we describe the baseline CPU-only performance.
We then measure the server-side performance in different testing configurations, in both standalone and realistic conditions. Finally, we scale up the workflow and make detailed measurements of performance. We also derive a scaling model for how we expect performance to scale and compare it to our measured results.

\vspace{0.5cm}
\subsection{CPU-only baseline}
\label{sec:cpuonly}

To compare against our heterogeneous computing system, we first measure the throughput of the CPU-only process. The workflow processes events from a Monte Carlo simulation of cosmic ray events in ProtoDUNE-SP, produced with the Corsika generator~\cite{corsikamc}. The radiological backgrounds, including $^{39}$Ar, $^{42}$Ar, $^{222}$Rn, and $^{85}$Kr, are also simulated using the \emph{RadioGen} module in LArSoft. Each event corresponds to a $3\unit{ms}$ readout window with a sampling rate of $0.5\unit{\ensuremath{\mu}s}$ per time tick. The total number of electronic channels is 15360. A white noise model with an RMS of $2.5\unit{ADC}$ was used. The workflows are executed on grid worker nodes running Scientific Linux 7, with CPU types shown in Table~\ref{tab:cputypes}. The fraction of all clients that ran with an average client-side batch size of $1693$ for each CPU type is shown in the second column of this table. Of these clients, 64\% of them ran on nodes with $10\unit{Gbps}$ network interface cards (NICs) and the remainder ran on nodes with $1\unit{Gbps}$ NICs. 

\begin{table}[tbh]
    \centering
    \begin{tabular}{lr}
         \hline\noalign{\smallskip}
         CPU type & \multicolumn{1}{c}{fraction (\%)} \\
         \noalign{\smallskip}\hline\noalign{\smallskip}
         AMD EPYC 7502 @ 2.5 GHz & 11.7 \\
         AMD Opteron 6134 @ 2.3 GHz & 0.6 \\
         AMD Opteron 6376 @ 2.3 GHz & 4.6 \\
         Intel Xeon E5-2650 v2 @ 2.6 GHz & 30.8 \\
         Intel Xeon E5-2650 v3 @ 2.3 GHz & 5.2 \\
         Intel Xeon E5-2670 v3 @ 2.3 GHz & 7.3 \\
         Intel Xeon E5-2680 v4 @ 2.4 GHz & 17.3 \\
         Intel Xeon Gold 6140 @ 2.3 GHz & 22.6 \\
         \noalign{\smallskip}\hline
    \end{tabular}
    \caption{CPU types and distribution for the grid worker nodes used for the ``big-batch'' clients (see text for more details).} 
    \label{tab:cputypes}
\end{table}

We measure the time it takes for each module in the reconstruction chain to run. We divide them into 2 basic categories: the non-ML modules and the ML module. The time values are given in Table~\ref{tab:cpu}.  
Of the CPU time in the ML module, we measure that $7\unit{s}$ is dedicated to data preprocessing to prepare for neural network inference, and the rest of the time, $213\unit{s}$, from the module is spent in inference. This is the baseline performance to which we will compare our results using GPUaaS.  It is important to know that further CPU optimization could improve performance, but would require drastic changes either to the workflows (multi-threaded ML inference in LArSoft) or to the software integration. In this study, we compare our approach against the ``out-of-the-box'' CPU implementation where the only addition on our end is a non-disruptive GPU service.  

\begin{table}[tbh]
    \centering
    \begin{tabular}{ccc}
         \hline\noalign{\smallskip}
         \multicolumn{3}{c}{Wall time (s)} \\
         ML module & non-ML modules & Total \\
         \noalign{\smallskip}\hline\noalign{\smallskip}
         220 & 110 & 330 \\ 
         \noalign{\smallskip}\hline
    \end{tabular}
    \caption{The average CPU-only wall time per job for the different module categories.}
    \label{tab:cpu}
\end{table}

\vspace{0.5cm}
\subsection{Server-side performance}
\label{sec:server}

To get a standardized measure of the performance, we first use standard tools for benchmarking the GPU performance.  Then we perform a stress test on our GPUaaS instance to understand the server-side performance under high load.

\vspace{1.0cm}
\paragraph*{Server standalone performance}

The baseline performance of the GPU server running the \emtkid model is measured using the \textit{perf\_client} tool included in the Nvidia Triton inference server package. The tool emulates a simple client by generating requests over a defined time period. It then returns the latency and throughput, repeating the test until the results are stable. We define the baseline performance as the throughput obtained the saturation point of the model on the GPU. We attain this by increasing the client-side request concurrency---the maximum number of unanswered requests by the client---until the throughput saturates. We find that the model reaches this limit quickly at a client-side concurrency of only 2 requests. At this point, the throughput is determined to be $20,000 \pm 2,000$ inferences per second. This corresponds to an event processing time of $2.7 \pm 0.3\unit{s}$. This is the baseline expectation of the performance of the GPU server.  

\vspace{1.0cm}
\paragraph*{Saturated server stress test}

To understand the behavior of the GPU server performance in a more realistic setup, we set up many simultaneous CPU processes to make inference requests to the GPU. This saturates the GPUs, keeping the pipeline of inference requests as full as possible. We measure several quantities from the GPU server in this scenario. To maximize throughput, we activate the dynamic batching feature of Triton, which allows the server to combine multiple requests together in order to take advantage of the efficient batch processing of the GPU. This requires only one line in the server configuration file.

\begin{figure}[tbh]
    \centering
    \includegraphics[width=0.55\textwidth]{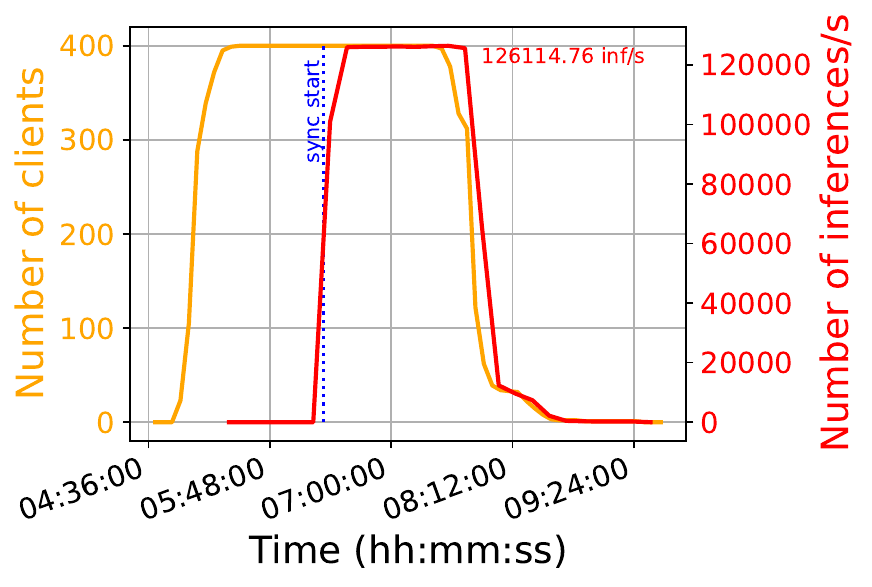}
    \includegraphics[width=0.44\textwidth]{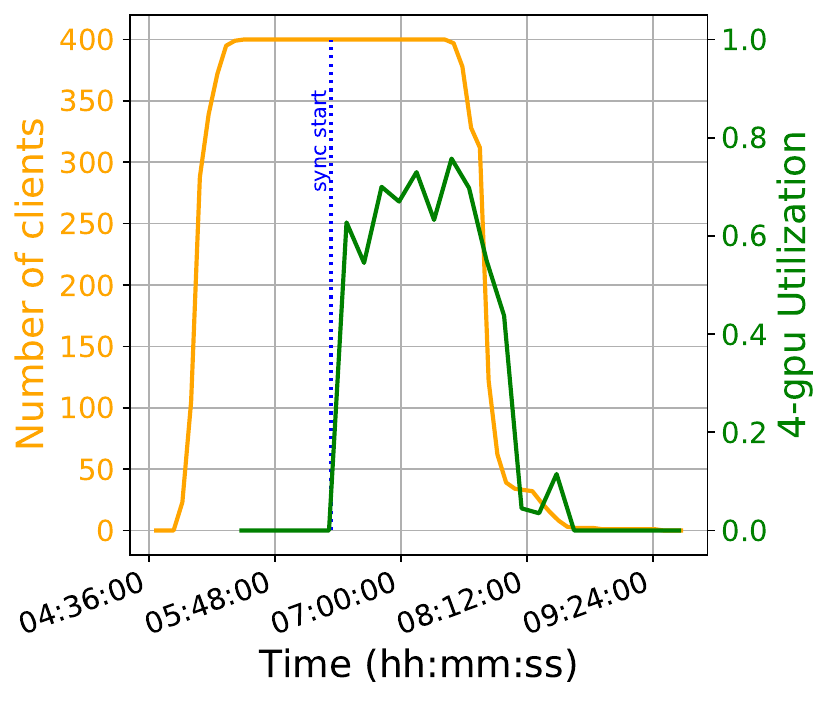}\\
    \includegraphics[width=0.5\textwidth]{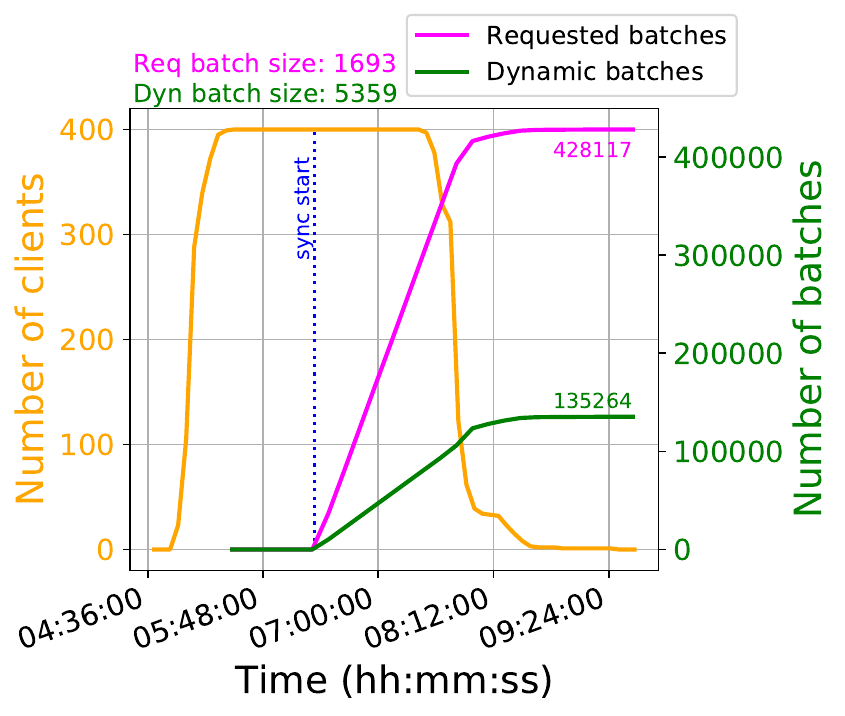}
    \caption{Top left: The number of inferences per second processed by the 4-GPU server, which saturates at approximately 126,000. Top right: The GPU usage, which peaks around 60\%. Bottom: The number of total batches processed by the 4-GPU server. The incoming batches are sent to the server with size 1693, but are combined up to size 5358 for optimal performance.}
    \label{fig:gpuinf}
\end{figure}

In this setup, we run 400 simultaneous CPU processes that send requests to the GPU inference server.  This is the same compute farm described in Sec.~\ref{sec:cpuonly}. The jobs are held in an idle state until all jobs are allocated CPU resources and all input files are transferred to local storage on the grid worker nodes, at which point the event processing begins simultaneously. This ensures that the GPU server is handling inference requests from all the CPU processes at the same time. This test uses a batch size of 1693. We monitor the following performance metrics of the GPU server in 10-minute intervals:
\begin{itemize}
    \item GPU server throughput: for the 4-GPU server, we measure that the server is performing about 122,000 inferences per second for large batch and dynamic batching; this amounts to 31,000 inferences per second per GPU. This is shown in Fig.~\ref{fig:gpuinf} (top left). This is higher than the measurement from the standalone performance client, by a factor of ${\sim}1.5$. For large batch and no dynamic batching, we observe similar throughput, while for small batch and no dynamic batching, we find that performance is a bit worse, close to the standalone client performance at 22,000 inf/s/GPU.  
    \item GPU processing usage: we monitor how occupied the GPU processing units are. We find that the GPU is ${\sim}60\%$ occupied during saturated processing. This is shown in Fig.~\ref{fig:gpuinf} (top right).
    \item GPU batch throughput: we measure how many batches of inferences are processed by the server. The batch size sent by the CPU processor is 1693 on average, but dynamic batching prefers to run at a typical batch size of 5358.  This is shown in Fig.~\ref{fig:gpuinf} (bottom).
\end{itemize}


\vspace{0.5cm}
\subsection{Scale out modeling}
\label{sec:model}

In the previous section, we discussed in detail the GPU server performance. With that information, we study in detail the performance of the entire system and the overall improvement expected in throughput.

To describe important elements of the as-a-service computing model, we first define some concepts and variables. Many of these elements have been described in previous sections, but we collect them here to present a complete model.

\begin{itemize}
    \item \tcpu is the total time for CPU-only inference as described in Sec.~\ref{sec:cpuonly}. This is measured to be $330\unit{s}$.
    \item $p$ is the fraction of \tcpu which can be accelerated, and conversely $1-p$ is the fraction of the processing that is not being accelerated. The ML inference task takes $220\unit{s}$, but we subtract \tpreproc, the time required for data preprocessing, which still occurs on the CPU even when the algorithm is offloaded to the GPU. This results in $p=0.65$.
    \item \tgpu is the time explicitly spent doing inference on the GPU. We measure this to be 22,000--31,000 inferences per second depending on whether or not dynamic batching is used. For 55,000 inferences per event, this turns out to be $1.77\unit{s}$ ($2.5\unit{s}$) when dynamic batching is enabled (disabled with small batch).
    \item \tpreproc is the time spent on the CPU for preprocessing to prepare the input data to be transmitted to the server in the correct format. This is measured to be $7\unit{s}$.
    \item \ttransmit is the latency incurred from transmitting the neural network input data. For 55,000 inferences per event, with each input a $48\times48$ image at $32\unit{bits}$, the total amount of data transmitted is about $4.1\unit{Gigabits}$ per event. Sec.~\ref{sec:k8s} specifies that each CPU process is allocated a $2\unit{Gbps}$ link on the server side while Sec.~\ref{sec:cpuonly} specifies $1\unit{Gbps}$ or $10\unit{Gbps}$ link speed on the client side. Therefore the communication bottleneck varies between $1\unit{Gbps}$ and $2\unit{Gbps}$ such that the total latency for transmitting data is between $2.05\unit{s}$ and $4.1\unit{s}$.
    \item \ttravel is the travel latency to go from the Fermilab data center, which hosts the CPU processes, to the GCP GPUs. This depends on the number of requests \nreq and the latency per request \treq. The latter can vary based on networking conditions, but we measure it to be roughly $12\unit{ms}$. The small batch size of 256 images requires $\nreq = 214$ to transmit the 55,000 images, while the large batch size of 1720 images requires $\nreq = 32$. Given these parameters, we find that the travel latency is $2.6\unit{s}$ ($0.38\unit{s}$) for small (large) batch size.
    \item $\tlatency = \tpreproc + \ttransmit + \ttravel$ summarizes the additional latency distinct from the actual GPU processing.
    \item \tideal is the total processing time assuming the GPU is always available; this is described in more detail in the next section. 
    \item \ncpu and \ngpu are the numbers of simultaneously running CPU and GPU processors, respectively.
\end{itemize}

With each element of the system latency now defined, we can model the performance of SONIC. Initially, we assume blocking modules and zero communication latency.
We define $p$ as the fraction of the event which can be accelerated, such that the total time of a CPU-only job is trivially defined as:
\begin{equation}
\centering
    \tcpu = (1-p) \times \tcpu + p \times \tcpu
\end{equation}
\noindent We replace the time for the accelerated module with the GPU latency terms:
\begin{equation}
    \tideal = (1-p) \times \tcpu + \tgpu + \tlatency.
\label{eq:ideal}
\end{equation}
\noindent This reflects the ideal scenario when the GPU is always available for the CPU job. We also include \tlatency, which accounts for the preprocessing, bandwidth, and travel time to the GPU. The value of \tgpu is fixed, unless the GPU is saturated with requests. We define this condition as how many GPU requests can be made while a single CPU is processing an event. The GPU saturation condition is therefore defined as: 
\begin{equation}
\frac{\ncpu}{\ngpu} > \frac{\tideal}{\tgpu}.
\end{equation}
Here, \tideal is equivalent to Eq.~\eqref{eq:ideal}, the processing time assuming there is no saturated GPU. There are two conditions, unsaturated and saturated GPU, which correspond to ${\frac{\ncpu}{\ngpu} < \frac{\tideal}{\tgpu}}$ and ${\frac{\ncpu}{\ngpu} > \frac{\tideal}{\tgpu}}$, respectively. We can compute the total latency (\tsonic) to account for both cases:
\begin{equation}
    \tsonic = (1-p) \times \tcpu + \tgpu\left[ 1+  \text{max}\left(0,\frac{\ncpu}{\ngpu} - \frac{\tideal}{\tgpu} \right) \right] + \tlatency.
    \label{eq:sat}
\end{equation}

\noindent Therefore, the total latency is constant when the GPUs are not saturated and increases linearly in the saturated case proportional to \tgpu. Substituting Eq.~\eqref{eq:ideal} for \tideal, the saturated case simplifies to:
\begin{equation}
    \tsonic = \tgpu \times \frac{\ncpu}{\ngpu}.
\end{equation}

\vspace{0.5cm}
\subsection{Measurements deploying SONIC}

To test the performance of the SONIC approach, we use the setup described in the ``server stress test'' in Section~\ref{sec:server}. We vary the number of simultaneous jobs from 1--400 CPU processes. To test different computing model configurations, we run the inference with two different batch sizes: 235 (small batch) and 1693 (large batch). This size is specified at run time through a parameter for the \emtkid module in the FHiCL~\cite{fhicl} configuration file describing the workflow.  With the small batch size, inferences are carried out in approximately 235 batches per event.  Increasing the batch size to 1693 reduces the number of inference calls sent to the Triton server to 32 batches per event, which decreases the travel latency. We also test the performance impact of enabling or disabling dynamic batching on the server.

\begin{figure}[tbh]
 \centering
 \includegraphics[width=0.48\textwidth]{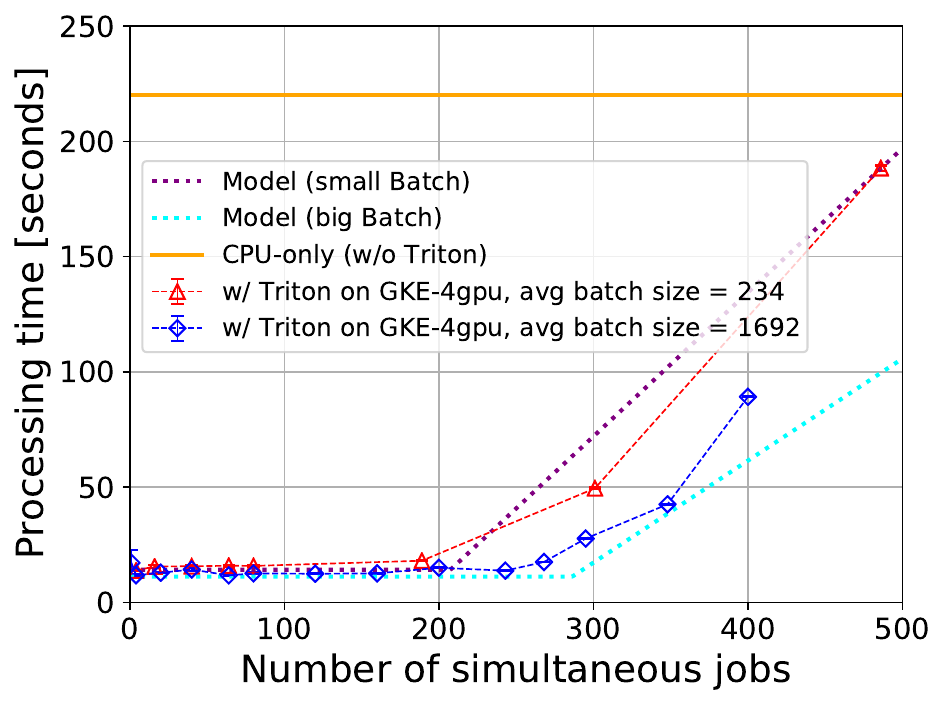}
 \includegraphics[width=0.48\textwidth]{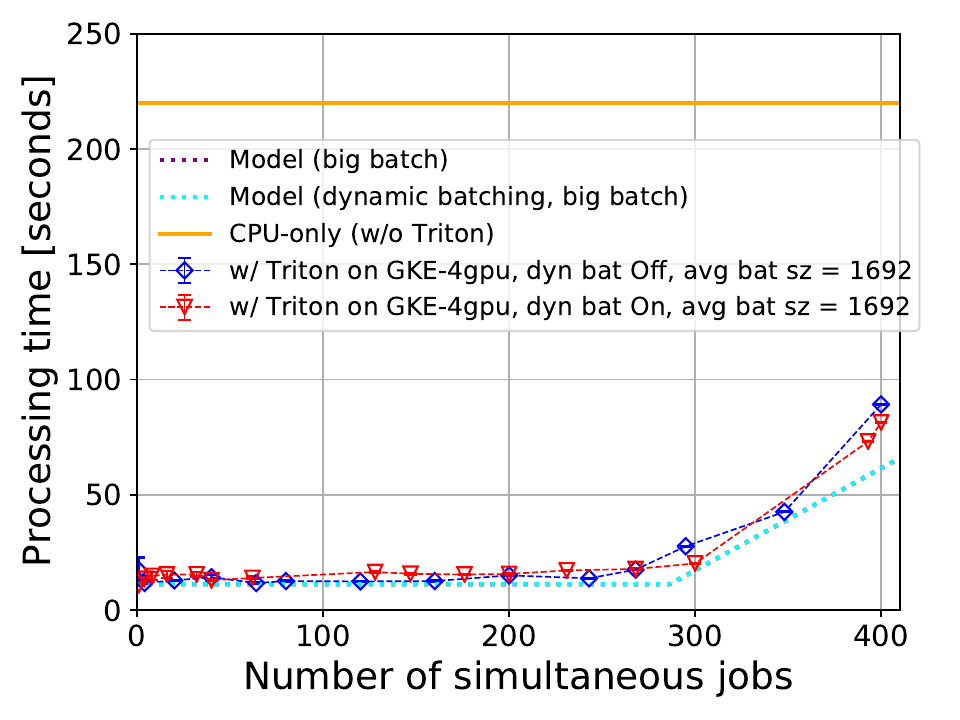}
 \caption{Processing time for the \emtkid module as a function of simultaneous CPU processes, using a Google Kubernetes 4-GPU cluster. Left: small batch size vs. large batch size, with dynamic batching turned off. Right: large batch size performance with dynamic batching turned on and off. In both plots, the dotted lines indicate the predictions of the latency model, specifically Eq.~\eqref{eq:sat}.}
 \label{fig:model_val}
\end{figure}

In Fig.~\ref{fig:model_val} (left), we show the performance results for the latency of the \emtkid module for small batch size vs. large batch size, with dynamic batching turned off. The most important performance feature is the basic trend. The processing time is flat as a function of the number of simultaneous CPU processes up to 190 (270) processes for small (large) batch size. After that, the processing time begins to grow, as the GPU server becomes saturated and additional latency is incurred while service requests are being queued. For example, in the large batch case, the performance of the \emtkid module is constant whether there are 1 or 270 simultaneous CPU processes making requests to the server. Therefore, using less than 270 simultaneous CPU processes for the 4-GPU server is an inefficient use of the GPU resources; and we find that the optimal ratio of CPU processes to a single GPU is 68:1.

As described in Section~\ref{sec:model}, $7\unit{s}$ of the module time is spent on the CPU for preprocessing to prepare the inputs for neural network inference. The term \ttravel is computed based on a measured round trip ping time of $12\unit{ms}$ for a single service request. Therefore, for small (large) batch size, the total \ttravel per event is $2.6\unit{s}$ ($0.38\unit{s}$). The difference between the corresponding processing times for the different batch sizes roughly corresponds to the $2.22\unit{s}$ difference in \ttravel times. We also see that in the small batch size configuration, the GPU server saturates earlier, at about 190 simultaneous CPU processes. In comparison, the large batch size server saturates at about 270 simultaneous processes. This is because the GPU is more efficient with larger batch size: at a batch size of 235 (1693), the GPU server can process about 80,000 (125,000) images per second. The overall performance using the SONIC approach is compared to the model from Section~\ref{sec:model}. We see that performance matches fairly well with our expectations.  

In Fig.~\ref{fig:model_val} (right), we show the performance of the SONIC approach for large batch size with dynamic batching enabled or disabled, considering up to 400 simultaneous CPU processes. We find that at large batch size, for our particular model, the large batch size of 1693 is already optimal and the performance is the same with or without dynamic batching. We also find that the model for large batch size matches the data well.

We stress that, until the GPU server is saturated, the \emtkid module now takes about $13\unit{s}$ per event in the most optimal configuration. This should be compared against the CPU-based inference, which takes $220\unit{s}$ on average. The \emtkid module is accelerated by a factor of 17, and the total event processing time goes from $330\unit{s}$ to $123\unit{s}$ on average, a factor of 2.7 reduction in the overall processing time.  The results are summarized in Table~\ref{tab:cpugpu}.

\begin{table}[tbh]
    \centering
    \begin{tabular}{cccc}
         \hline\noalign{\smallskip}
         \multicolumn{4}{c}{Wall time (s)} \\
         & ML module & non-ML modules & Total \\
         \noalign{\smallskip}\hline\noalign{\smallskip}
         CPU only & 220 & 110 & 330 \\ 
         CPU + GPUaaS & 13 & 110 & 123 \\ 
         \noalign{\smallskip}\hline
    \end{tabular}
    \caption{A comparison of results in Table~\ref{tab:cpu} with results using GPUaaS}
    \label{tab:cpugpu}
\end{table}

\vspace{0.5cm}
\subsection{Server-side performance}
\label{sec:server}

Finally, it is important to note that throughout our studies using commercially available cloud computing, we have observed that there are variations in the GPU performance. This could result from a number of factors beyond our control, related to how CPU and GPU resources are allocated and configured in the cloud. Often, these factors are not even exposed to the users and therefore difficult to monitor. That said, the GPU performance, i.e. the number of inferences per second, is a non-dominant contribution to the total job latency. Volatility in the GPU throughput primarily affects the ratio of CPU processes to GPUs. We observe variations at the 30\%--40\% level, and in this study, we generally present conservative performance numbers.

\section{Summary and Outlook}
\label{sec:outlook}
In this study, we demonstrate for the first time the power of accessing GPUs as a service with the Services for Optimized Network Inference on Coprocessors (SONIC) approach to accelerate computing for neutrino experiments. We integrate the Nvidia Triton inference server into the LArSoft software framework, which is used for event processing and reconstruction in liquid argon neutrino experiments. We explore the specific example of the ProtoDUNE-SP reconstruction workflow. The reconstruction processing time is dominated by the \emtkid module, which runs neural network inference for a fairly traditional convolutional neural network algorithm over thousands of patches of the ProtoDUNE-SP detector. In the standard CPU-only workflow, the module consumes 65\% of the overall CPU processing time.  

We explore the SONIC approach, which abstracts the neural network inference as a web service. A 4-GPU server is deployed using the Nvidia Triton inference server, which includes powerful features such as load balancing and dynamic batching. The inference server is orchestrated using Google Cloud Platform's Kubernetes Engine. The SONIC approach provides flexibility in dynamically scaling the GPUaaS to match the inference requests from the CPUs, right-sizing the heterogeneous resources for optimal usage of computing. It also provides flexibility in dealing with different machine learning (ML) software frameworks and tool flows, which are constantly improving and changing, as well as flexibility in the heterogeneous computing hardware itself, such that different GPUs, FPGAs, or other coprocessors could be deployed together to accelerate neural network algorithms. 
In this setup, the \emtkid module is accelerated by a factor of 17, and the total event processing time goes from $330\unit{s}$ to $123\unit{s}$ on average resulting in a factor of 2.7 reduction in the overall processing time. We find that the optimal ratio of GPUs to simultaneous CPU processes is 1 to 68.

With these promising results, there are a number of interesting directions for further studies.
\begin{itemize}
    \item \textbf{Integration into full-scale production}: A natural next step is to deploy this workflow at full scale, moving from 400 simultaneous CPU processes up to 1000--2000. While this should be fairly straightforward, there will be other interesting operational challenges to be able to run multiple production campaigns. For example, the ability to instantiate the server as needed from the client side would be preferable. The GPU resources should scale in an automated way when they become saturated. There are also operational challenges to ensure the right model is being served and server-side metadata is preserved automatically.
    \item \textbf{Server platforms}: Related to the point above, physics experiments would ultimately prefer to run the servers without relying on the cloud, instead using local servers in lab and university data centers. Preliminary tests have been conducted with a single GPU server at the Fermilab Feynman Computing Center. Larger-scale tests are necessary, including the use of cluster orchestration platforms. Finally, a similar setup should be explored at high performance computing (HPC) centers, where a large amount of GPU resources may be available.
    \item \textbf{Further GPU optimization}: Thus far, the studies have not explored significant optimization of the actual GPU operations. In this paper, a standard 32-bit floating point implementation of the model was loaded in the Triton inference server. A simple extension would be to try model optimization using 8-bit or 16-bit operations. This would further improve the GPU performance and thereby increase the optimal CPU-to-GPU ratio. More involved training-side optimizations might yield similar physics performance at a reduced computational cost. For example, quantization-aware training tools such as QKeras~\cite{Coelho:2020zfu} and Brevitas~\cite{alessandro_pappalardo_2020_3979501} could maintain performance at reduced precision better than simple post-training quantization.  
    \item \textbf{More types of heterogeneous hardware}: In this study, we have deployed GPUs as a service, while in other studies, FPGAs and ASICs as a service were also explored. For this particular model and use case, with large batch sizes, GPUs already perform very well. However, the inference for other ML algorithms may be more optimal on different types of heterogeneous computing hardware. Therefore, it is important to study our workflow for other platforms and types of hardware.
\end{itemize}

By capitalizing on the synergy of ML and parallel computing technology, we have introduced SONIC, a non-disruptive computing model that provides accelerated heterogeneous computing with coprocessors, to neutrino physics computing. We have demonstrated large speed improvements in the ProtoDUNE-SP reconstruction workflow  and anticipate more applications across neutrino physics and high energy physics more broadly.

\section*{Acknowledgments}
We acknowledge the Fast Machine Learning collective as an open community of multi-domain experts and collaborators. This community was important for the development of this project. We acknowledge the DUNE collaboration for providing the ProtoDUNE-SP reconstruction code and simulation samples. We would like to thank Tom Gibbs and Geetika Gupta from Nvidia for their support in this project.  We thank Andrew Chappell, Javier Duarte, Steven Timm for their detailed feedback on the manuscript.

M. A. F., B. Ha., B. Ho., K. K., K. P., N. T., M. W., and T. Y. are supported by Fermi Research Alliance, LLC under Contract No. DE-AC02-07CH11359 with the U.S. Department of Energy, Office of Science, Office of High Energy Physics.
N. T. and B. Ha. are partially supported by the U.S. Department of Energy Early Career Award.  
K. P. is partially supported by the High Velocity Artificial Intelligence grant as part of the Department of Energy High Energy Physics Computational HEP sessions program.
P.~H. is supported by NSF grants \#1934700,  \#193146., J.~K. is supported by NSF grant \#190444. 
Cloud credits for this study were provided by Internet2 managed Exploring Cloud to accelerate Science (NSF grant \#190444). 
\clearpage
\bibliographystyle{JHEP} 
\bibliography{main}

\end{document}